\documentclass[apj]{emulateapj}
\usepackage{graphicx}
\usepackage{amsmath}

\received{September 13, 2023} 
\revised{December, 10,, 2023} 
\accepted{December, 14 2023} 

\usepackage[colorlinks,
        hypertexnames=false,
        pagebackref=false,
        linkcolor=blue,    
    citecolor=blue,        
    filecolor=magenta,     
    urlcolor=blue          
]{hyperref}                

\newcommand{\degree}{$^\circ$}

\newcommand{\kms}{km s$^{-1}$} 

\shorttitle{Filament eruption from AR 13283 and intense geomagnetic storm on 23 April 2023}
\shortauthors{Vemareddy}

\begin{document}
\title{Filament eruption from active region 13283 leading to fast halo-CME and intense geomagnetic storm on 23 April 2023}


\author{P.~Vemareddy}    
\affil{Indian Institute of Astrophysics, II Block, Koramangala, Bengaluru-560 034, India}

\begin{abstract}
Using multi-instrument and multi-wavelength observations, we studied a CME eruption that led to intense geomagnetic storm on 23~April~2023. The eruption occurred on April~21 in solar active region 13283 near the disk-center. The AR was in its decay stage, with fragmented polarities and a pre-existing long filament channel, a few days before the eruption. The study of magnetic field evolution suggests that the flux-rope (filament) has been built up by monotonous helicity accumulation over several days, and further converging and canceling fluxes lead to helicity injection change resulting in the unstable nature of the magnetic flux-rope (MFR) and its further eruption. Importantly, the CME morphology revealed that the MFR apex underwent a rotation of upto 56\degree~in clockwise-direction owing to its positive helicity. The CME decelerates in the LASCO-FOV and has a plane-of-sky velocity of 1226~km/s at 20\,R$_\odot$. In the Heliospheric Imager FOV, the CME lateral expansion is tracked more than the earthward motion. This implies that the arrival time estimation is difficult to assess. The in-situ arrival of ICME shock was at 07:30~UT on April 23, and a geomagnetic storm commenced at 08:30\,UT. The flux rope fitting to the in-situ magnetic field observations reveals that the MC flux rope orientation is consistent with its near Sun orientation, which has a strong negative Bz-component. The analysis of this study indicates that the near-Sun rotation of the filament during its eruption to the CME is the key to the negative Bz-component and consequently the intense geomagnetic storm.
\end{abstract}

\section{Introduction}
\label{Intro}
Coronal mass ejections (CMEs) are large scale magnetized plasma structures originating from the Sun. They propagate outward through interplanetary space, affecting the planetary atmospheres (e.g., \citealt{Webb2012_lrsp}). Their propagation toward Earth was established to be the cause of the most severe geomagnetic disturbances on Earth {e.g., \citealt{Gosling1991_GeoMag_act, Gosling1993_FlareMyth, webb2000,ZhangJie2007_GeostCME}. Therefore, it is of scientific and technological interest to understand the complete picture of CMEs, including their magnetic structure and the mechanisms involved in their origin from the source regions, evolution, and propagation from the Sun to the Earth \citep{ZhangJie2021_EarthEff_SolTrans}.

Historically, CMEs have been observed in the images of white-light Coronagraphs onboard OSO7 \citep{MacQueen1974}, Skylab \citep{Sheeley1980_Solwind}, and Solar Maximum Mission (SMM; \citealt{House1981}). When a CME happens, coronagraphs use Thomson-scattered light from free electrons in coronal and heliospheric plasma to observe the outward flow of density structures coming from the Sun. Following these early space coronagraphs, LASCO coronagraph aboard SoHO, launched in 1995, is still continuously capturing white-light observations of the Sun. From 2006, the coronagraph instruments onboard the twin STEREO \citep{howard2008} spacecrafts started providing vantage point white-light images in a much wider FOV, enabling one to track the CME propagation up to 330 R\,$_\odot$. The CMEs in white-light images are often observed in association with filament or prominence structures (e.g., \citealt{Webb1987_ActCMEProm, Gopalswamy2003_PromEru, Vemareddy2012_FilErup, Vemareddy2017_PromEru}), X-ray sigmoids \citep{canfield1999,moore2001,Vasantharaju2019_FormEruSig}, and EUV hot channels (e.g., \citealt{zhangj2012,ChengX2013, vemareddy2014_IniErup_11719, Vemareddy2022_hotchan}) from the solar disk. Therefore, the white light observations of the CMEs are frequently accompanied by eruptions from the solar disk, e.g., by GONG H$\alpha$ telescope, the SOHO Extreme Ultraviolet Imaging Telescope, the STEREO Extreme-UltraViolet Imager (EUVI), and the recently launched SDO/AIA instrument. The in situ counterparts of the CMEs are called interplanetary coronal mass ejections (ICMEs), whose observations at 1 au, and behind are supplied by the spacecraft instruments onboard, e.g., Voyagers, Ulysses, Helios, Wind, ACE, and STEREO etc.

The CMEs have a traditional three-part structure, which is usually understood to be compressed plasma in front of a flux rope, a cavity, and a bright filament or prominence surrounding the cavity \citep{illing1985}. Therefore, the CME magnetic configuration is a magnetic flux rope (MFR) with a helical-field wound around the central axis. After ejection, the CME largely maintains its magnetic configuration or topology, as a twisted MFR and therefore is able to continuously propagate outward through the heliosphere interacting with the ambient solar wind. A statistical study by \citet{Vourlidas2013} suggests that at least 40\% of CMEs observed by space-borne instruments have a clear MFR structure. According to many studies based on satellite and ground-based data, this MFR structure in the outer corona is thought to be an evolved form of the magnetic structure of filaments observed in H$\alpha$ or sigmoidal structures in soft X-rays in the source regions.

\begin{figure*}[!ht]
\centering
\includegraphics[width=.99\textwidth,clip=]{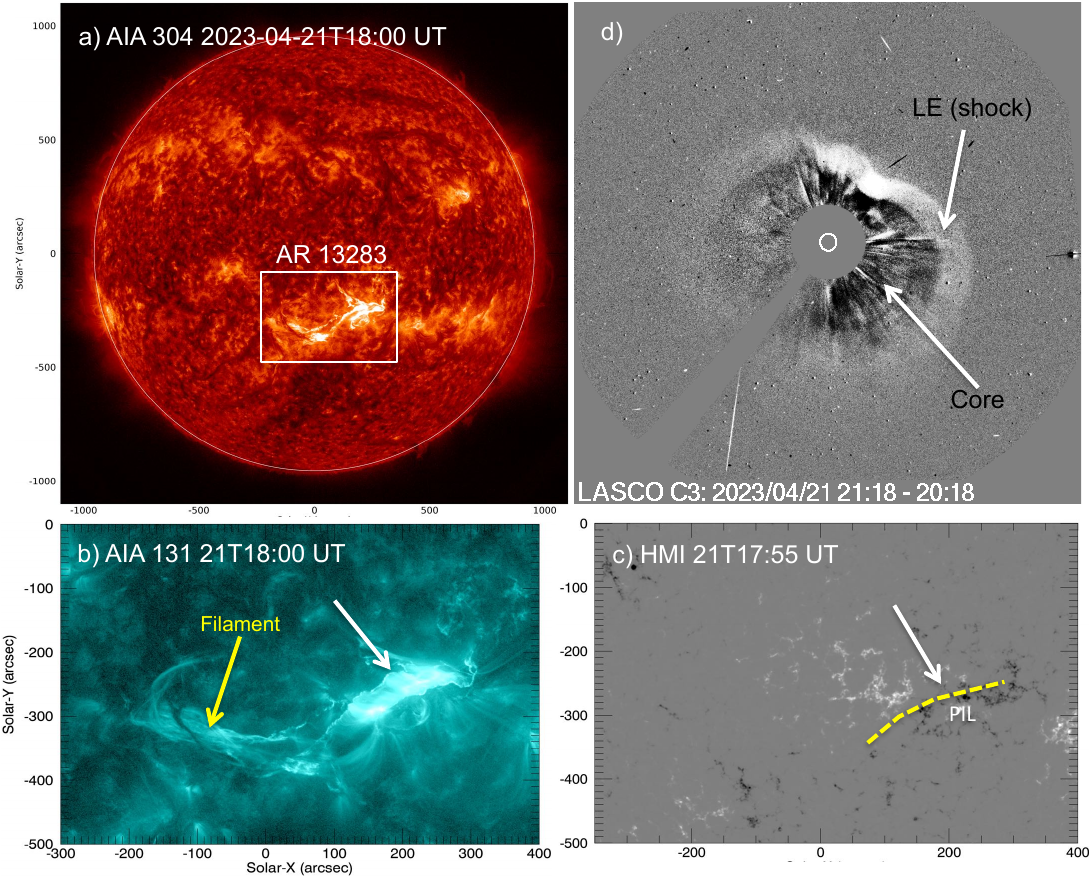}
\caption{The CME eruption from the Sun occurred on April 21, 2023, at 18:00 UT. {\bf a)} AIA 304~\AA~image of the Sun showing the location of the AR 13283 (inset) as the source region of the CME eruption. {\bf b)} AIA 131\AA images of the AR during the eruption. {\bf c)} HMI magnetic field observations of the AR 13283. The yellow dashed curve traces the polarity inversion line between fragmented opposite magnetic polarities. The long filament channel remains intact, and the erupting feature is a manifestation of the sheared magnetic loops in the vicinity of the PIL. {\bf d)} LASCO/C3 running difference image delineates the Earth-directed halo CME with bright leading edge and core. }
\label{fig1}
\end{figure*} 

Geoeffective CMEs are those that are directed toward the Earth and are called Halo CMEs, therefore, the source region should be near the visible disk center \citep{SrivastavaN2004_GeoSt, ZhangJie2007_GeostCME}. These front-side halo CMEs associated with strong soft X-ray flares tend to be the most geoeffective \citep{Gopalswamy2007_GeoEff_halo}. After ejection from the inner corona, the CMEs evolve with distinct kinematic, geometric, and magentic properties, which determine their geoeffectiveness. The kinematic studies reveal that the CMEs evolve with slow, fast, and gradual acceleration in the inner corona upto 0.1 au \citep{zhangj2004}. Its further propagation in the outer heliosphere, is governed solely by the interaction of the CME and the ambient solar wind via aerodynamic drag \citep{cargill2004,vrsnak2010} which causes the faster CMEs decelerate and slower ones to accelerate, tending to equalise their speeds with the speed of the ambient solar wind. Most importantly, the CME is geoeffective when its interplanetary magnetic field is southward-pointing, which depends on the orientation of the MFR with respect to the ecliptic. Some CMEs are observed to rotate in the corona due to a variety of mechanisms. Simulations showed the rotation is due to kink instability, and the direction of the rotation depends on the handedness of the flux rope magnetic field \citep{torok2003,lynch2009}. CMEs have also been observed to deflect both in latitude and longitude  \citep{Isavnin2013_CME_1au}. Depending on the relative positions of the coronal features,  the deflection motion was frequently described as being toward the heliospheric current sheet \citep{cremades2004} or away from coronal holes \citep{Gopalswamy2009_CME_deflect}. These geometric changes of the CMEs in the corona are the key for explaining the CME impact on Earth. For example, \citet{KayC2017_CMEDef_11158} studied the trajectory of the series of CMEs launched from AR 11158, located near the disk center, and found that only one CME encountered the Earth. The other missing CMEs were found to deflect north up to 30\degree~near the Sun. 

\begin{figure*}[!ht]
\centering
\includegraphics[width=.9\textwidth,clip=]{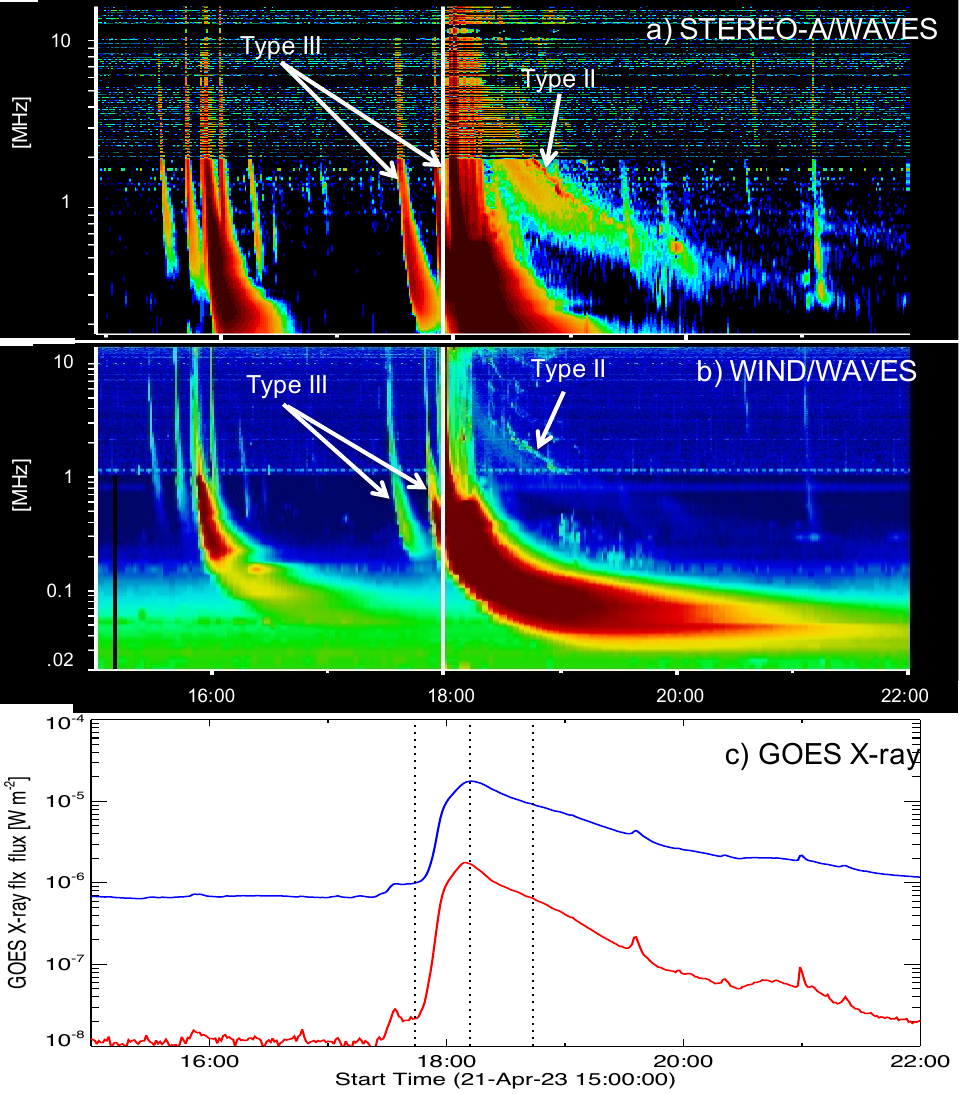}
\caption{Emission signatures of the CME eruption from the AR 13283.
	{\bf (a)} Radio dynamic spectrum obtained from WAVES onboard STA. Arrows point to the type II, III radio burst triggered by the eruption. 
	{\bf (b)} Radio dynamic spectrum obtained by WIND/WAVES instrument located at the Lagrange L1 point. Arrows refer to the type III bursts originating from the CME associated flare. 
	{\bf (c)} Disk integrated GOES X-ray flux. The peak value of the flux refer to M1.7 class flare occurred from AR 13283. Vertical dotted lines refer to flare start (17:44 UT), peak (18:12 UT) and end (18:44 UT) times respectively. }
 \label{fig_radio}
\end{figure*} 

In the interest of space-weather impact, identifying the geomagnetic source regions is very important to understand the characteristics of the magnetic structure responsible for the CME eruption and its evolution in the sun-earth line. In the 24th solar cycle, only three strong storms occurred with Dst index $<-$175 nT, i.e., on 17 March 2018, 23 June 2015 and 26 August 2018 \citep{Gopalswamy2022_WhatIsUnusual}. In this study, we study the CME source region of the intense storm occurred on April 23, 2023 with Dst Index -212 nT. The source region is identified as active region 13283. An overview of the observations is given in section~\ref{ObsOver}. Results of the analysis on magnetic field evolution in the AR, CME onset mechanism, CME propagation, and in situ observations are described in Section~\ref{Res}. A summary of the results with a discussion is provided in section~\ref{SummDisc}.

\begin{figure*}[!ht]
\centering
\includegraphics[width=.75\textwidth,clip=]{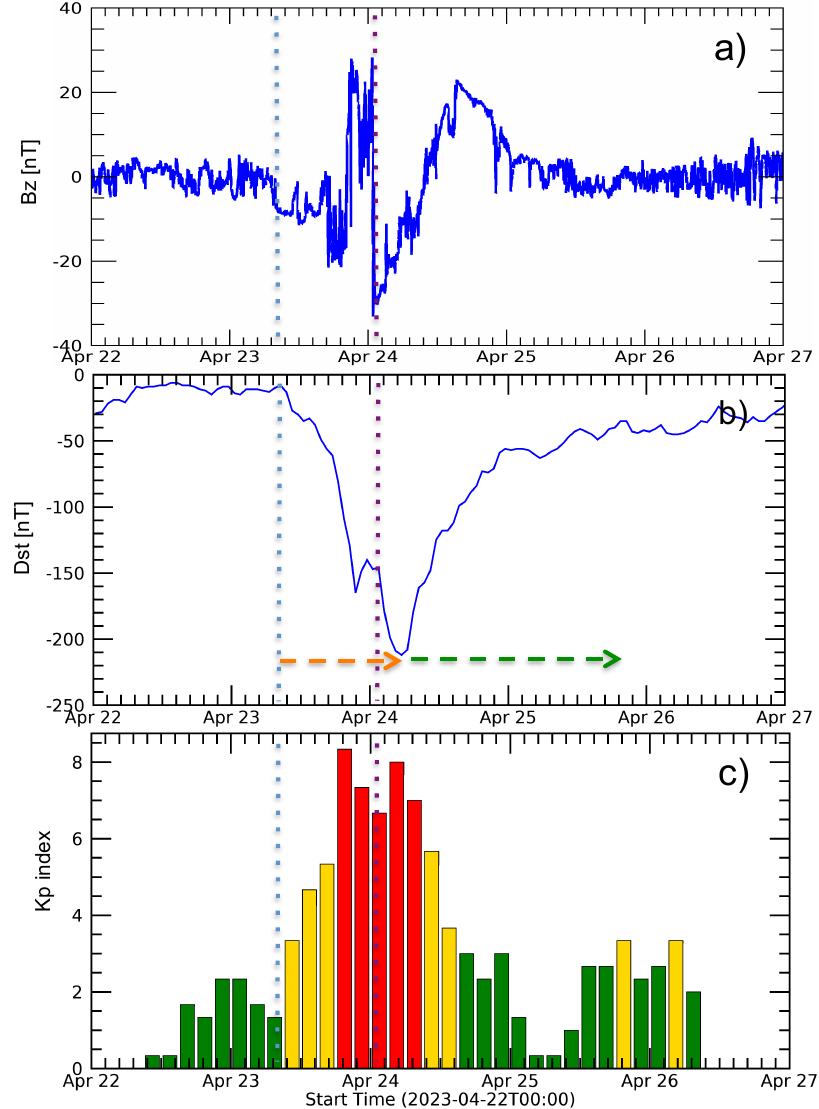}
\caption{In-situ observations of the CME encounter the Earth's atmosphere. {\bf a)} the Bz component of the magnetic field with time. The Bz started becoming negative at around 07:30 UT on April 23. Blue, and magenta dotted vertical lines correspond to shock (23/07:30 UT), and MC (24/01:00 UT) arrival times. {\bf b)} Dst index with time. The Dst index started decreasing to negative values around 08:30 UT on April 23. It peaked at -212 nT at 05:30 UT on April 24, 2023, indicating an intense and long-duration geomagnetic storm. The storm's main phase (orange dashed arrow) persists for 22 hrs, followed by a long recovery phase for about a couple of days. {\bf c)} Histogram of Kp index values. Red, yellow, and blue bars refer to index values 0-3, 3-6, 6-9, respectively. The Kp index values reach up to eight during the storm's main phase, indicating an intense geomagnetic storm of G4 severity as per space weather classification.   }
\label{fig_kpi}
\end{figure*}

\begin{figure*}[!ht]
\centering
\includegraphics[width=.99\textwidth,clip=]{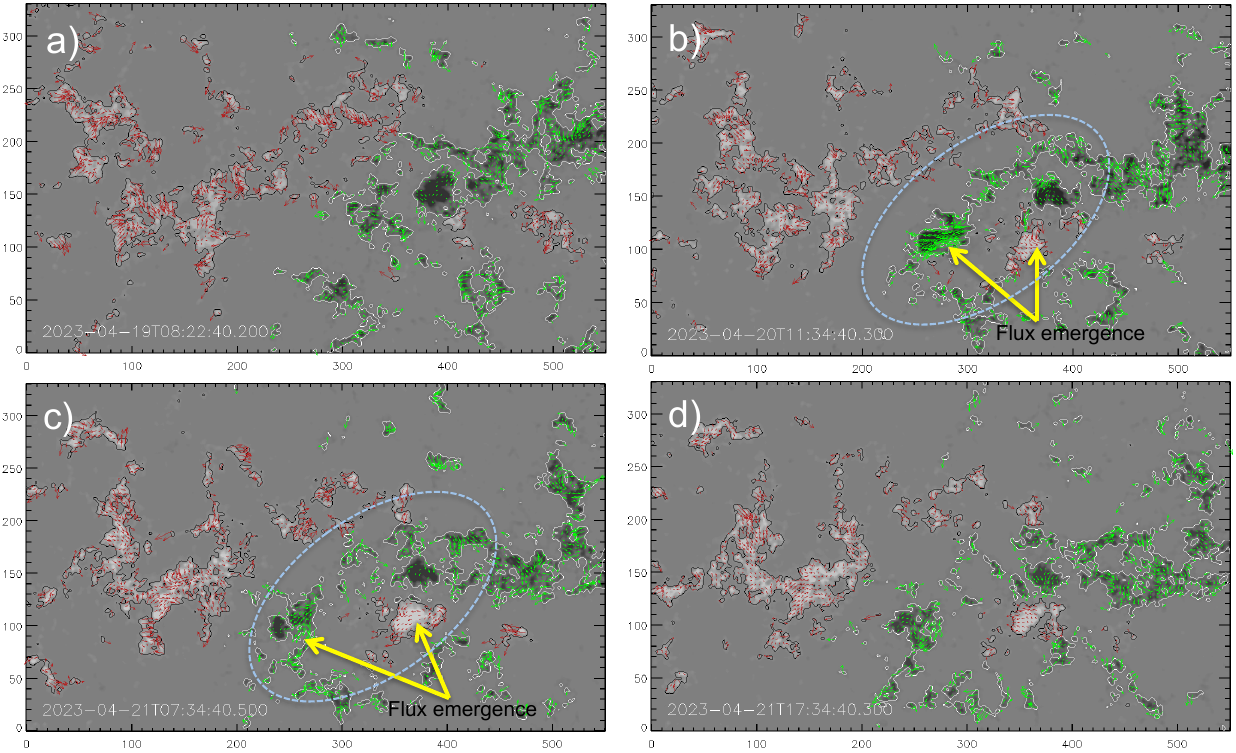}
\caption{Horizontal velocity of flux motions in the AR 13283. Background image is $B_z$ map with contours $\pm$150 G. The arrows indicate the direction of motion, and their length is proportional to the magnitude of a maximum of 0.7 km\,s$^{-1}$. Large shear motions of the fluxes can be noticed in the region enclosed by blue ovals. Also, fluxes emerge and move apart during the evolution. In all panels, axis units are in pixels of 0.5".} 
\label{fig_vel}
\end{figure*} 

\section{Overview of Observations}
\label{ObsOver}
We study the event of a large-scale CME eruption launched from the source AR 13283 on April 21, 2023. We use multi-spacecraft and multi-wavelength observations to study magnetic characteristics of the source region, the initiation of the CME from the source region, and its propagation toward the Earth. 

The source region observations are well covered by the instruments onboard the Solar Dynamics Observatory (SDO). Atmospheric Imaging Assembly \citep{Lemen2012} observes the full disk of the Sun continuously in seven extreme ultraviolet (EUV) channels. These channels correspond to the chromosphere in the temperature range of around 20,000 K to the corona at 10 million K. The cadence of the observations is 12s at a pixel scale of 0.6 arcsec. The corresponding magnetic field measurements at the photosphere are obtained from the Helioseismic and Magnetic Imager at a pixel size of 0.5 arcsec. Figure~\ref{fig1}a displays the full disk image of the Sun in AIA 304~\AA~waveband. The eruption's onset (CME launch) occurred at around 18:00 UT from AR 13283, located near the disc center (S21\degree W11\degree) with a long filament channel present along its polarity inversion line (PIL, yellow dashed curve). The observations of the AR in AIA 131~\AA~waveband and the corresponding magnetic field measurements at the photosphere are shown in Figure~\ref{fig1}b-c. The AR consists of fragmented polarities without a major sunspot, which suggests that the AR is in the decaying phase of its evolution. The solar feature of the explosion manifests from the shearing and converging motion of the fluxes along the PIL. In fact, a long filament channel preexisted along the PIL of the opposite magnetic fluxes. The eruption feature is a filament as part of this multiply-threaded channel. However, the left part of the long filament channel appeared unaffected during the eruption.  (Ref Figure~\ref{fig_aia_fil}). 

The white-light observations of the CME are obtained from Large Angle Spectrometric Coronagraph \citep[LASCO;][]{Brueckner1995} on board the Solar and Heliospheric Observatory (SOHO) as well as from the Sun Earth Connection Coronal and Heliospheric Investigation (SECCHI;  \citealt{howard2008}) onboard the Solar-Terrestrial and Relational Observatory (STEREO; \citealt{Kaiser2008}). LASCO provides white light images in the field-of-view (FOV) of 1.5-6 R$_\odot$ (C2), and 3-32 R$_\odot$ (C3) to study the CME dynamics in the near-Sun corona. Heliospheric Imagers onboard SECCHI image the sun in a much wider FOV (HI1: 15-90 R$_\odot$ and HI2: 70-330 R$_\odot$ ) and they are suited to understand the CME propagation beyond the Earth. At the time of this CME launch, the STEREO-A (STA) has a separation angle of 10\degree~with respect to the Earth, so the CME is viewed with a slightly similar morphology as with LASCO. Figure~\ref{fig1}d presents the running difference LASCO/C3 image showing the large scale halo CME from the AR. The linear speed of the CME projected in the plane of sky is 1284 \kms which classifies the CME as fast. CMEs with speeds greater than 960 km/s and widths greater than 66\degree~are connected with type II radio bursts and are highly geoeffective \citep{gopalswamy2001}. 

\begin{figure*}[!ht]
\centering
\includegraphics[width=.99\textwidth,clip=]{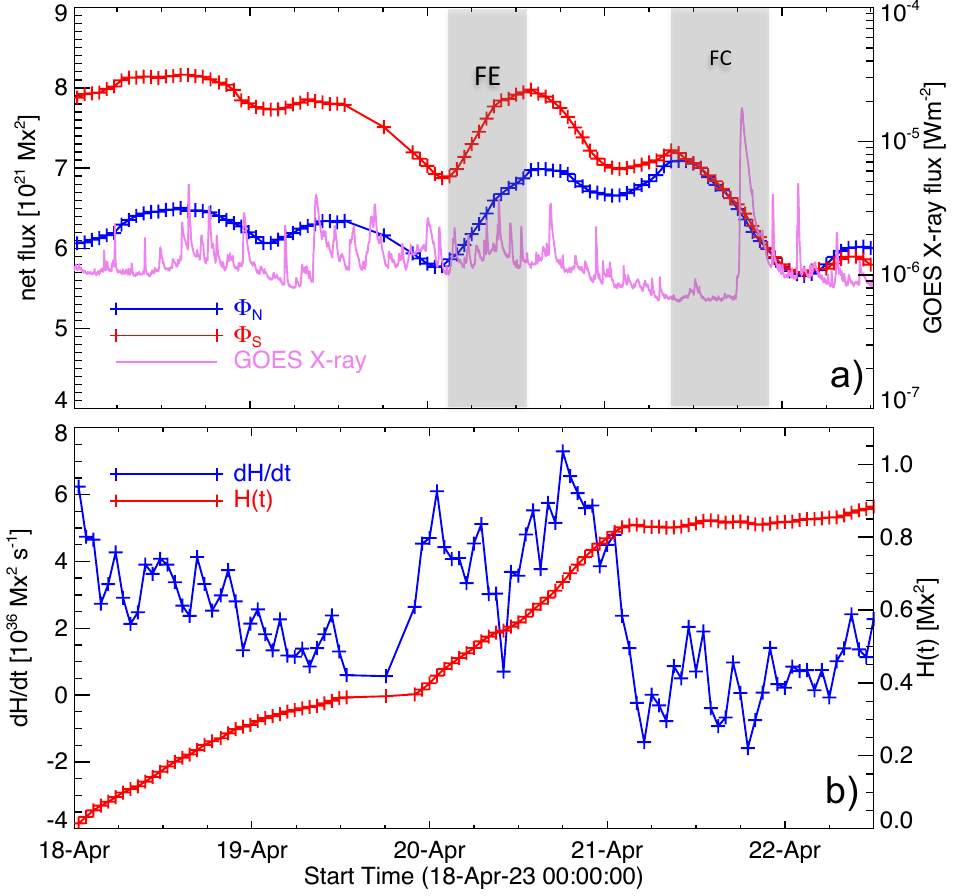}
\caption{Magnetic evolution in AR 13283. {\bf a)} Net flux in positive (north) and negative (south) magnetic polarities. Disk-integrated GOES X-ray (1.0–8.0~\AA~ passband) flux is also shown with the y-axis scale on the right. Grey bands refer to the phase of the flux emergence (FE) and flux cancellations (FC) by converging motions. Note that the M1.7 flare associated with the fast CME occurred during the converging and canceling of flux motions. {\bf b)} the helicity injection rate ($dH/dt$) with time. Accumulated helicity, $H(t)$, is also plotted with a y-axis scale on the right side.  }
\label{fig_hinj}
\end{figure*} 

The CME is associated with a GOES class M1.7 flare, as indicated by the disk-integrated soft X-ray flux plotted in Figure~\ref{fig_radio}c. The start, peak, and end times are 17:44 UT, 18:12 UT, and 18:44 UT, respectively. The radio dynamic spectrum obtained from the Radio and Plasma Wave Experiment (WAVES, \citealt{Bougeret1995}) onboard the Wind spacecraft detects the launch of type III radio bursts co-temporal with flare start and peak times (Figure~\ref{fig_radio}b). In addition to type III, the STA/WAVES dynamic spectrum detects type II radio bursts, which arise from shocks driven by CME \citep{Cane1987_IPShocks, Gopalswamy2019_TypeII_catalog}. The shock front is seen to persist up to 20:30 UT, as evidenced in LASCO white-light images, correspondingly, the type II emissions prolong.

The WIND satellite also records magnetic and plasma observations at the L1 point, which are obtained at a cadence of 92 seconds. The data demonstrate a distinct in situ counterpart of the CME (ICME) on April 23, with a total field intensity above 30 nT with a rotating magnetic field component (Ref Figure~\ref{fig_mc}). Figure~\ref{fig_kpi}a depicts the observations of the z-component of the magnetic field ($B_z$) with respect to time. The shock appears to have hit at 07:30 UT on April 23,  with the $B_z$ component becoming increasingly negative (southward $B_z$ component). With these observations of southward $B_z$, the Disturbance-storm-time (Dst) index values\footnote{obtained from \url{https://wdc.kugi.kyoto-u.ac.jp/dst\_realtime/202304/index.html}}, as shown in Figure~\ref{fig_kpi}b, indicate the start of a geomagnetic storm. The Dst profile is well correlated with the $B_z$ component associated with the shock-sheath, and magnetic cloud regions. The storm peaked at -212 nT around 05:30 UT on April 24. This storm has a Kp index values of up to 8 and is classified as G4 severe. It is the first major storm in solar cycle 25 to cause stunning northern lights (auroras) to extend to lower latitudes around the world.

\begin{figure*}[!ht]
\centering
\includegraphics[width=.99\textwidth,clip=]{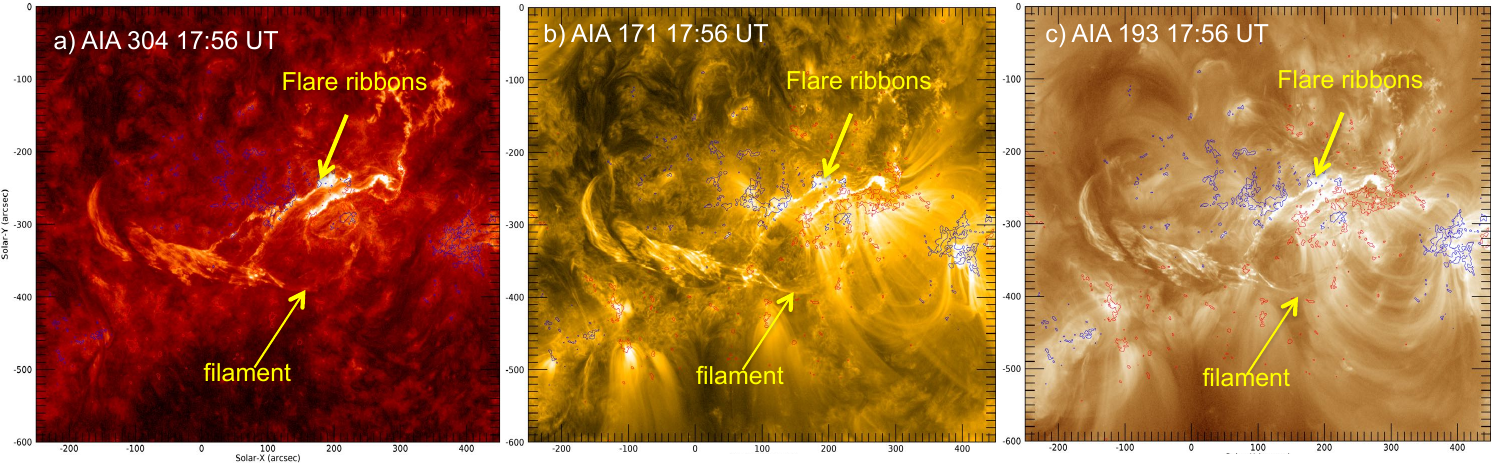}
\caption{EUV images during the onset of the eruption. {\bf a)} AIA 304~\AA~Image showing the filament section being on rise motion. Flare ribbons are formed on each side of the PIL. {\bf b-c)} AIA 171 and 193~\AA~images show the erupting filament. In all panels, contours of Bz ($\pm$150 G) are overlaid. These images reveal the filament threads wound in right-hand direction. }
\label{fig_aia_fil}
\end{figure*}

\begin{figure*}
\centering
\includegraphics[width=.95\textwidth,clip=]{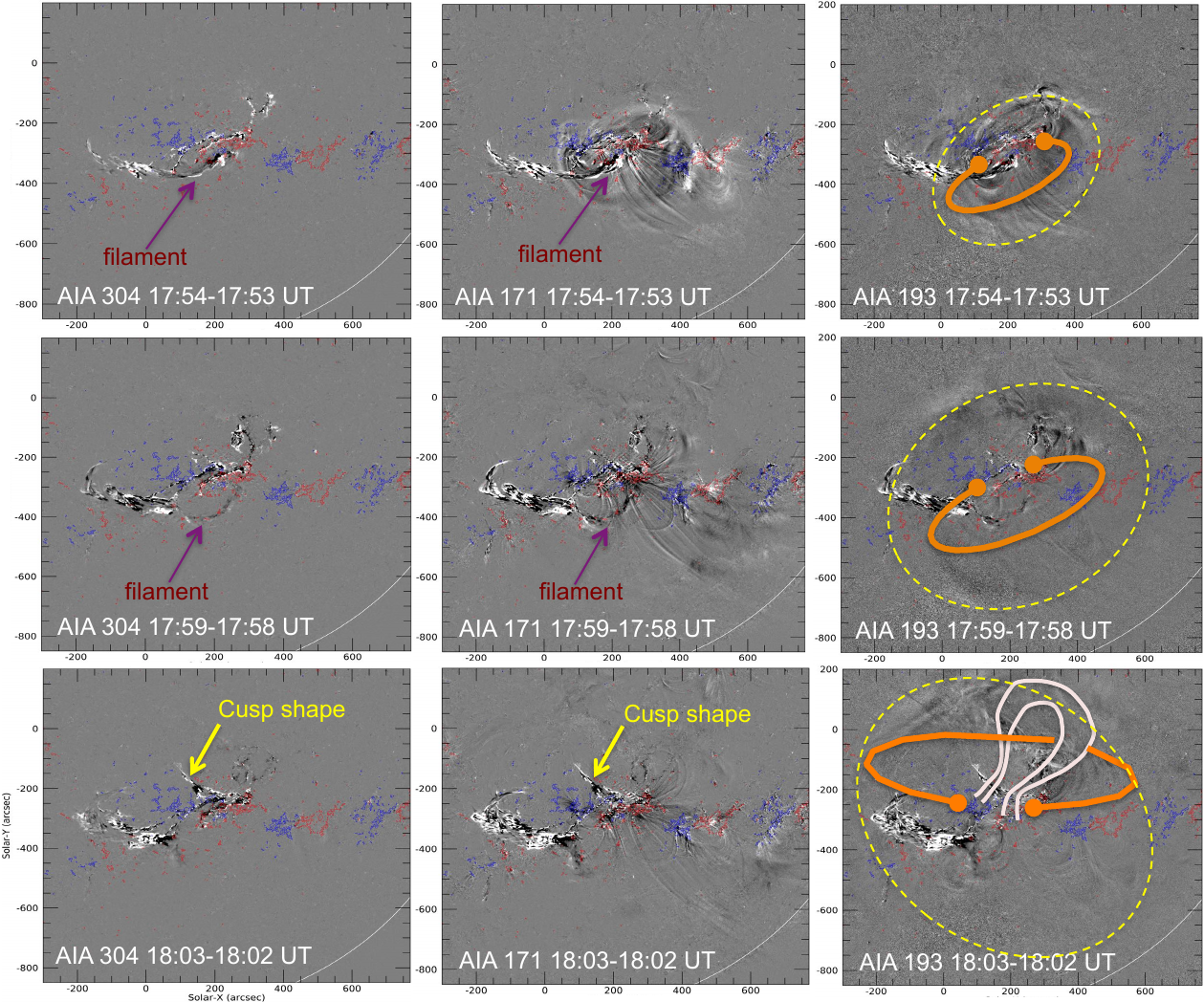}
\caption{Onset of the eruption from AR 13283. {\bf left column:} AIA 304 \AA~difference images showing the rising filament section at different times. A cusp shape is formed underneath the rising filament after 18:02 UT. {\bf middle column:} AIA 171 \AA difference images showing the rising filament and the oval shape disturbance due to rising filament as a blob-like structure. {\bf right column:} AIA 193~\AA~difference images show the oval-shaped disturbance surrounding the rising filament environment. The eruption scenario is illustrated in the panels, with the orange color curve representing the erupting filament and the overlying loops undergoing reconnection as pink curves. All panels are overlaid with Bz contours at $\pm$150 G. To better understand the eruption onset, the image sequence from 17:51-18:04 UT is attached as a video. (An animation of this figure is available.) }
\label{fig_rdif}
\end{figure*}

\begin{figure*}[!ht]
\centering
\includegraphics[width=.9\textwidth,clip=]{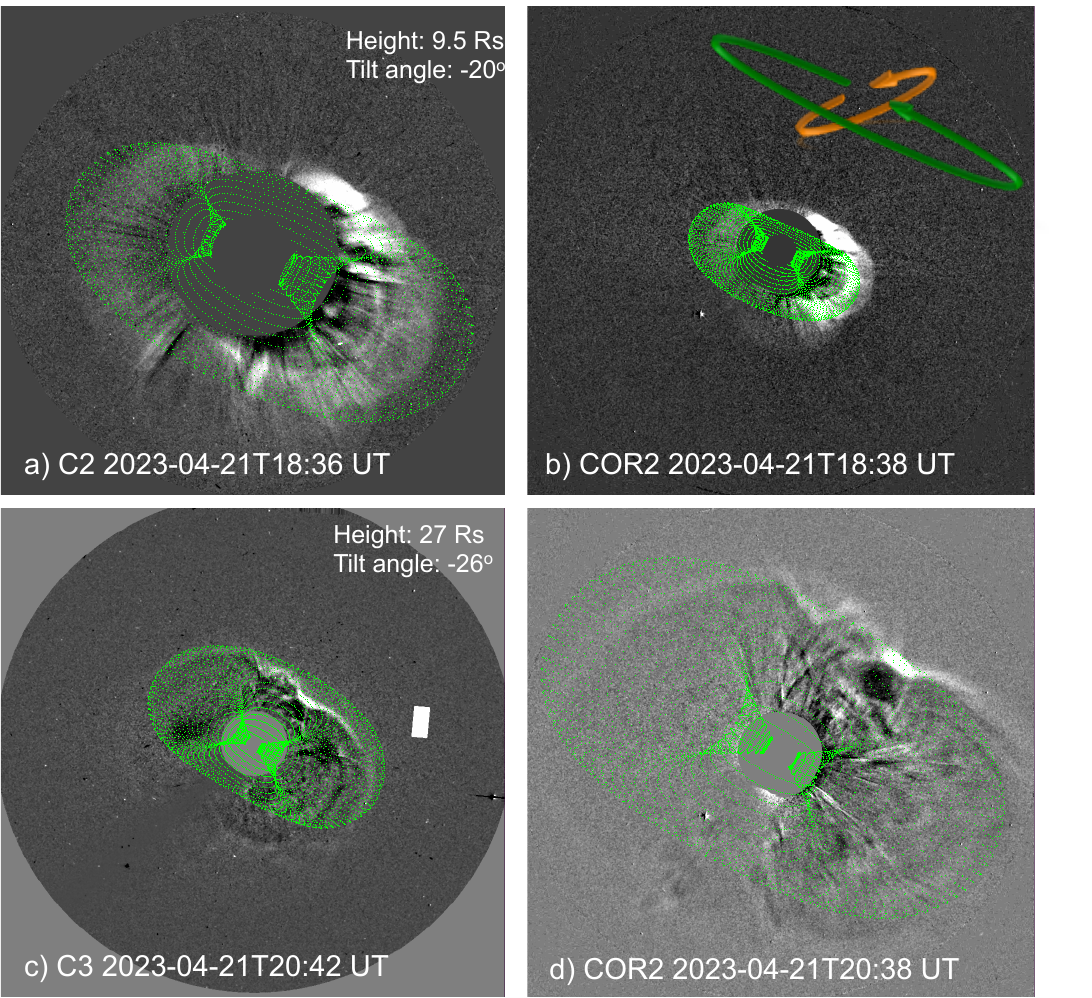}
\caption{GCS fit to CME morphology observed from COR2 and LASCO. {\bf a-b)} CME morphology observed from C2 and COR2, at 18:36 UT and 18:38 UT respectively. MFR {\bf axis} orientation is indicated as seen in the source region (orange) and LASCO (green) {\bf c-d)} CME morphology observed from C3 and COR2, at 20:36 UT and 20:38 UT respectively. In all the panels, the best-fitted wireframe of the MFR in the GCS model is overplotted. Since the STA and SoHO are separated at 10\degree, the CME morphology is similar in both views. Importantly, the MFR oriented from north-east to south-west direction with a tilt angle of -20\degree~at 18:36 UT and -26\degree~at 20:36 UT. Compared to source region PIL, the filament apex appears to rotated by about 56\degree~during its eruption to CME within 10\,R$_\odot$}
\label{fig_gcs}
\end{figure*}

\begin{figure*}
\centering
\includegraphics[width=.99\textwidth,clip=]{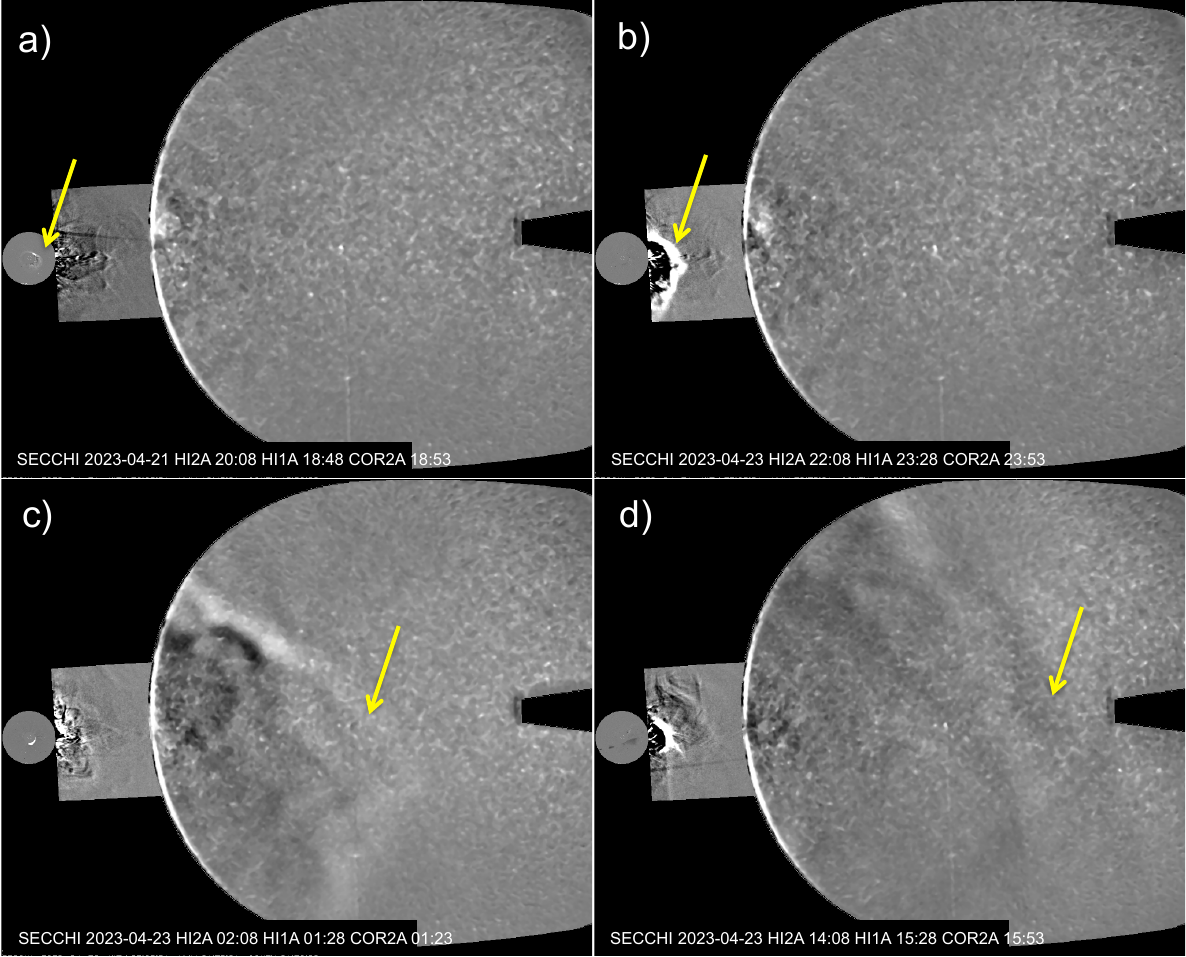}
\caption{CME propagation from low corona into the heliosphere.  {\bf a-d)} Combined running difference images of STA/COR2 (2.5-15 R$_\odot$), STA/HI1 and STA/HI2 during April 21–23, 2023. The yellow arrow points to the leading edge of the CME at different epochs of its propagation. Note that since the CME is propagating towards the STA spacecraft, the propagation in the images mostly represents the lateral expansion of the CME. The sequence of images displayed in each panel is attached as a video stream. (An animation of this figure is available).}
\label{fig_sta}
\end{figure*}

\section{Results}
\label{Res}
\subsection{Magnetic Evolution of the AR}
To study the magnetic field evolution in the AR, we obtain time series (every 12 minutes) of line-of-site magnetic field observations from the HMI. The magnetic field measurements at each instant of time are projected to the disk center by the cylindrical equal area projection method to minimize the projection effects. Then we choose the area of AR extent such that the foot points of the loops connected to the filament channel are covered over the evolution time under study. From these magnetograms, we derive the horizontal velocity of flux motions by using the Differential Affine Velocity Estimator (DAVE, \citealt{schuck2005}) method. The velocity field is displayed in Figure~\ref{fig_vel}a, where horizontal velocities (arrows) are plotted on Bz images. Contours of 150 G are also overlaid to identify the boundaries of individual polarity regions. The velocity field is spread up to a maximum value of 0.8 \kms indicating that different features move at different velocities. We found that emerging flux regions moved apart after their emergence.

The time evolution of the net magnetic flux is plotted in Figure~\ref{fig_hinj}a. The net flux in the north (south) polarity is around $6\times10^{21}$ Mx ($8\times10^{21}$ Mx). These fluxes are half of the values typically noticed in the growing ARs consisting of large-scale sunspot polarities \citep{vemareddy2019_VeryFast, Vemareddy2017_OppoHel}. The net flux profile increases during the first half of April 20 corresponding to the flux emergence, and 12 hours before the eruption, the net flux decreases by 1$\times$10$^{21}$ Mx in each polarity. From the velocity and magnetic fields, we also calculated the helicity injection rate ($dH/dt$) to evaluate the net helicity coming from the flux motions \citep{Demoulin2003_MagEneHel,vemareddy2012_hinj,Vemareddy2017_OppoHel}. The $dH/dt$ and $H$ with time are plotted in Figure~\ref{fig_hinj}b. The $dH/dt$ profile evolves with positive values around a mean value of $3.5\times10^{36}$ Mx$^2\,s^{-1}$ till April 21, after which it becomes marginally negative and then oscillates around zero values. The flux emergence phase is seen to pump additional helicity flux above the mean value. A positive sign of helicity refers to the right-hand helicity of the magnetic structure, which is also in agreement with the windings of the filament threads (Figure~\ref{fig_aia_fil}). During the four days of evolution, the flux motions accumulate a coronal helicity of $0.9\times10^{42}\,Mx^2$. It is worth pointing out that the filament channel was present well before April 18, therefore the magnetic structure exists with pre-accumulated helicity in the corona. Accounting for the net average flux in the AR as $7\times10^{21}$ Mx, the average twist of the flux tube ($H/\Phi^2$) could be $0.02-0.1 $ turns, which is in the range of several erupting ARs \citep{vemareddy2019_VeryFast}. In conclusion, the large shear and converging motions build the twisted flux about the PIL, which was set to erupt during the phase when flux motions started injecting weak helicity flux with alternate polarity into the AR corona. 

\subsection{Initiation and onset of the CME eruption }
The magnetic evolution in the AR is favorable for the eruption of twisted fields. Before the four days of the eruption, a long filament channel already existed, lying above the PIL. Its one leg lies in the leading negative polarity, and the other lies on the extreme east edge of the following positive polarity. Figure~\ref{fig_aia_fil} displays the EUV images showing the filament and the surrounding loops. These images reveal the orientation of the filament axis and the PIL, which are found in a south-east to north-west direction at a tilt angle of 30\degree. The filament threads, which are wound in the right-hand direction, are clearly visible along the filament channel. The winding direction reflects the right-handed helicalness or positive helicity of the magnetic field, in agreement with the sign of the helicity injection estimates. The overall shape of the filament is a forward S-shape, which indicates positive helicity and is compatible with the hemispheric rule of sigmoids being in the south hemisphere. 

About two hours before the main eruption, the filament channel shows signs of eruption, like brightening along its length, threads being reorganized, and the surrounding loop structures being heightened. The initiation of the eruption started at around 17:42 UT, with the sub-filament section along the PIL starting to rise.  This means that the large-scale filament is a combination of sub-twisted loop systems originating from different locations along the PIL.  With this rise, the GOES X-ray flux also increases, commencing the flare at 17:44 UT. Figure 5 delineates the rising filament section and the two flare ribbons at 17:56 UT. In order to visualize the filament and the surrounding loop system, we analyze running difference images, as displayed in Figure~\ref{fig_rdif}. AIA 304~\AA~difference images show clearly the rising filament section with time and its disappearance after 18:03 UT, whereas AIA 171 and 193 difference images give the impression of a blob-like structure expanding, which forms an oval shape in exactly the same direction as the PIL, i.e., south-east to north-west. Further rise motion brought the filament into the phase of the onset of its eruption, and this was clearly associated with the formation of the cusp-shaped loops underneath the expanding filament. The start of the filament rise motion at 17:42 UT and its onset at 18:03 UT fall into the scenario of tether-cutting reconnection \citep{moore2001}. The reconnection of the surrounding less sheared loops adds more field to the filament, which is regarded as the MFR, and the main phase of the eruption occurs as a result of the run-away tether cutting reconnection, which is linked to strong flare ribbons at the peak flare time. This eruption scenario is illustrated in the right panels of Figure~\ref{fig_rdif}, with the orange color curve representing the erupting filament and the overlying loops undergoing reconnection as pink curves. With the benefit of vantage point observations, such eruptions are being investigated more thoroughly, particularly the kinematics of slow and fast rising motion \citep{Vemareddy2012_FilErup, Vemareddy2022_hotchan}.

\subsection{Near sun rotation of the MFR}
To determine the underlying large-scale structure and orientation of the MFR, we exploited the plane of sky-projected coronagraph images from different vantage points. For this purpose, we employed the Graduated Cylindrical Shell (GCS; \citealt{thernisien2009}) model. In this model, the large-scale structure of the MFR is approximated either by the conical legs or the curved (tubular) fronts. The model is based on height-aspect ratio ($\kappa$), half-angle ($\alpha$), tilt angle ($\gamma$), latitude ($\theta $) and longitude ($\phi$). The underlying assumption of this model is that the pre-eruptive magnetic configuration of the source AR constrains the magnetic orientation of an erupting MFR. This is based on an analysis of several CMEs and the magnetic configurations of their source regions \citep{cremades2004}, which is incorrect in a number of observed cases.

In Figures~\ref{fig_gcs}, we displayed the difference images of STA/COR2, LASCO/C2, and C3, obtained at two different epochs. The wire frame (dotted green) from the GCS fit is overlaid in these panels. Since SoHO and STA spacecrafts are separated by 10\degree, the morphology of the CME is observed to be similar from the two different vantage points. The GCS fit to the CME morphology is implemented with a GUI procedure \texttt{rtsccguicloud.pro} available in the STEREO tree of SSWIDL software \citep{freeland1998}. While fitting the GCS model, the tilt angle of the erupting feature and the PIL orientation in the source region are generally considered because the erupting feature (filament or prominence) is assumed to be the MFR. However, we notice that the PIL orientation (south-east to north-west) in the source region is quite different from the CME morphology, which is in a north-east to south-west direction. It appears that the MFR (erupting feature) from the source region has undergone a rotation of about 60\degree. Therefore, we use this information to set the tilt angle by visually matching the observed CME morphology. From this fit, the tilt angle is found to vary from -20\degree~to -25\degree~while the CME evolves from 18:36 UT to 20:36 UT. Also, the corresponding heights are noted as 10 R$_\odot$ and 27 R$_\odot$ respectively. From these two heights in time, we found that the CME velocity in 3D space exceeds 1500 \kms. 

\begin{figure*}
\centering
\includegraphics[width=.8\textwidth,clip=]{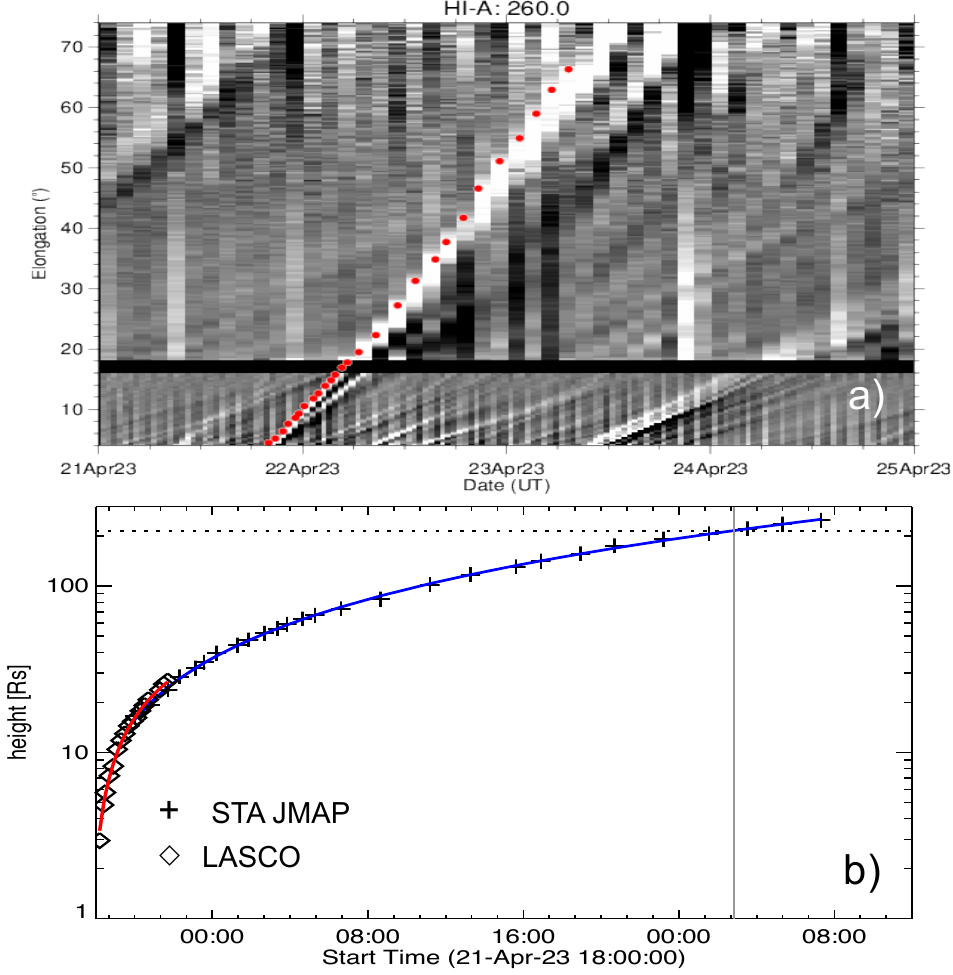}
\caption{CME kinematics in the plane-of-sky observations. a) A time-elongation map (J-map) constructed from the slices at a position angle of 260\degree. Red dots are the trace of the leading edge of the propagating CME. b) Height-time data of the CME. The red curve is a second order fit to the LASCO data (``$\diamond$" symbol), and the blue curve is a second order fit to J-map data (``+" symbol). The Sun-Earth distance is shown with horizontal dotted line, and vertical grey line represents the time (23/02:50 UT) of Sun-Earth distance passage in the plane-of-sky. }
\label{fig_ht}
\end{figure*}

\begin{figure*}[!ht]
\centering
\includegraphics[width=.9\textwidth,clip=]{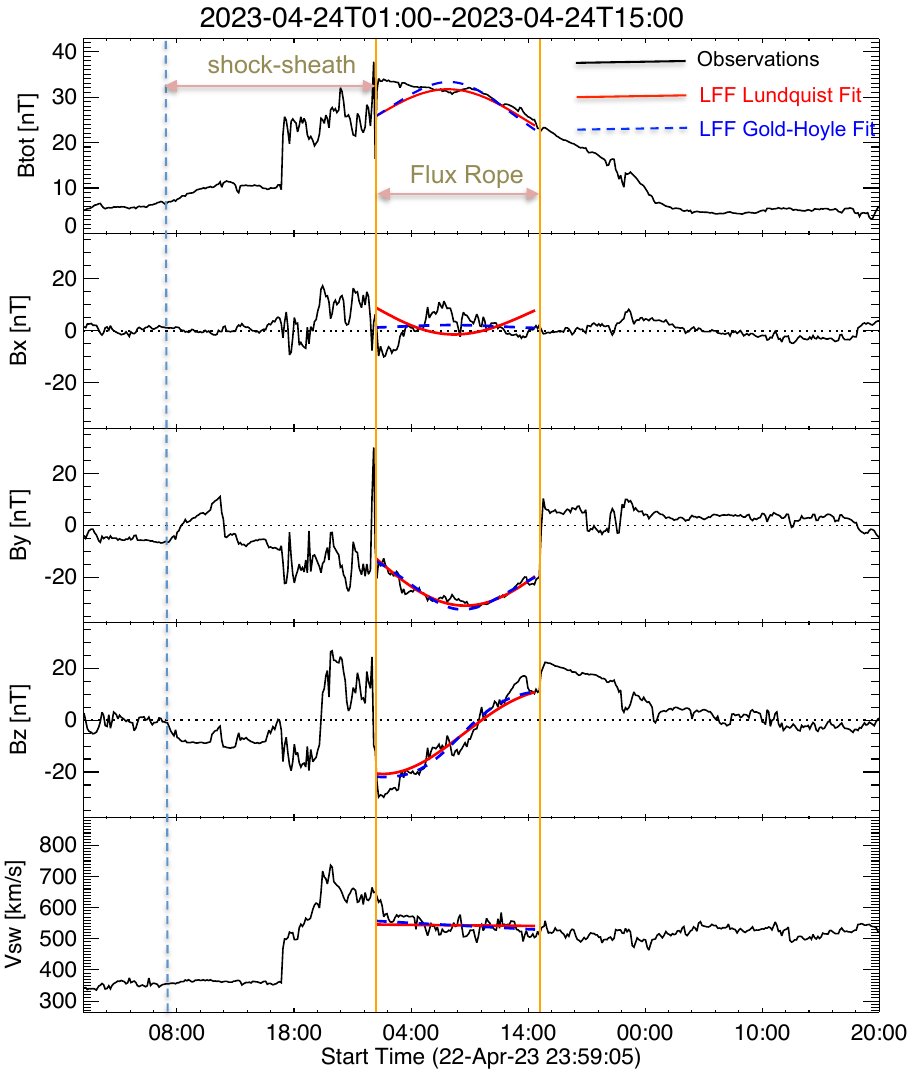}
\caption{In-situ observations of the CME (ICME) from WIND spacecraft. Panels from top to bottom show plots of magnetic field strength ($B_{\rm tot}$) and its components ($B_x$, $B_y$, $B_z$), and solar wind speed ($V_{\rm sw}$) as a function of time, respectively. The blue dashed vertical line marks the arrival time of shock (23/07:30 UT) at the spacecraft, and the orange vertical lines correspond to the MC start and end times. The blue and red curves are the fitting results of the Gold-Hoyle and Lundquist models, respectively. The fitting yields MC orientations of -5\degree~and -15\degree~latitude respectively, indicating a mostly southward Bz component.  }
\label{fig_mc}
\end{figure*}

\subsection{Propagation of the CME}
After the CME’s first emergence in LASCO C2 fov at 18:12 UT, its propagation in the C3 and COR2 is very clearly observed in the running difference images. Because of line-of-sight propagation, the CME motion in these images only represents lateral expansion. The CME further propagation is tracked in the wide-angle observations of the HI1 and HI2 of STA. The running different images of individual instruments are combined to make a composite running difference image at each instant of time so that the CME’s heliospheric propagation is visualized continuously while it transits from one fov to the other upto 370 R$_\odot$. In Figure~\ref{fig_sta}, we display these composite running images at four epochs. The bright shock front is evidently noticed, along with the CME leading edge in HI2. However, it is difficult to identify the precise boundaries of the CME core and leading edge. The morphology is not spherically symmetric rather a cone shape inclined roughly about 25\degree~to the ecliptic plane, consistent with the near Sun CME orientation. 

The event is also cataloged in the HELCATS list\footnote{\url{https://www.helcats-fp7.eu/catalogues/event\_page.html?id=HCME\_A\_\_20230421\_02}}, providing useful kinematic information from the analysis of time-elongation map (J-map) constructed from combined difference images \citep{sheeley1999,davies2009} as displayed in Figure~\ref{fig_ht}a. The time-elongation, in terms of height, is plotted in Figure~\ref{fig_ht}b. Also, we plot the height-time data obtained from the LASCO CME catalogue\footnote{\url{https://cdaw.gsfc.nasa.gov/CME\_list/UNIVERSAL\_ver1/2023\_04/univ2023\_04.html}}. A second order polynomial function best fits these observed height-time data of CME propagation. Starting from a velocity of 1459 km/s, the CME decelerates at a rate of 14 m\,s$^{-2}$ in LASCO fov. At a height of 20 R$_\odot$, the velocity of the CME is 1226 km/s. In the HI fov, the travel time of 1 au distance corresponds to 33 hours (3:00 UT on April 23) from CME onset at 18:00 UT on April 21. A difference of 4.5 hours compared to the observed arrival time of the ICME shock is due to projection effects. This emphasizes the difficulty of tracking the earthward propagation of the CME in this case.

Since the CME launched at 20:30 UT on April 21 at the Sun, the time of arrival of the shock front is 35 hours. As a geomagnetic storm commences at the same time as a shock front's arrival, it is important to assess the arrival time of the CME frontal structure, which is ICME shock front. We use an initial CME speed at 20 R$_\odot$, in the drag based model. Also, the drag coefficient $\gamma=0.15\times10^{-7}$ km$^{-1}$ and a background solar wind speed of 400 km/s are used. For the observed 1226 km/s speed, the transit time exactly matches with the arrival of the ICME sheath. Any higher value of drag coefficient may require a higher value of CME speed to match with its observed in situ arrival.

\subsection{In situ magnetic field observations}
The WIND observations of the in-situ magnetic field are plotted in Figure~\ref{fig_mc}. These observations are one minute averaged in the Geocentric Solar Ecliptic (GSE) coordinate system. The shock front encounters the spacecraft at 07:30 UT on April 23. Following the ICME shock-sheath region, the MC passes, which is identified by the systematic variation of the magnetic field components, indicating a strong magnetic field (30 nT) associated with an MFR-like structure. The Bz component is rotating, with its sign changing from negative to positive. Also, the solar wind velocity increases from the background values of around 360 \kms. Based on these observations, we identified the time intervals for the MC as 01:00-15:00 UT on April 24.

The MC signatures are often modelled by an MFR. We employ the \citet{Lundquist1950} and \citet{goldhoyle1960} (GH) models of cylindrically symmetric linear force-free magnetic fields. Because the MC grows during transit, the fitting technique takes this into consideration \citep{vemareddy2016_MagCld}. The magnetic field profile along the spacecraft's observational path in the Lundquist model is determined by the orientation of the flux rope axis, i.e., elevation and azimuth angle ($\theta$ and  $\phi$), the closest approach of the observational path p, the flux rope diameter D, the helicity sign H, and the field strength $B_0$ at the flux rope's centre \citep{lepping1990}. In the GH model, the twist parameter $T_0$ is added to these parameters. The helicity sign is decided on a trial basis; in this case, it is positive and consistent with the helicity sign of the source region, and the rest of the parameters are determined when the fitting converges in the least squares sense. As a measure of goodness of the fit, we compute 
the root-mean-square deviation ($rms=\sqrt{\sum_{i=1,N}\left ( \mathbf{B}_o(t_i)-\mathbf{B}_m(t_i) \right )^2/N} $)  between the observed magnetic field ($\mathbf{B}_o(t_i)$) and the modeled field ($\mathbf{B}_m(t_i)$) \citep{marubashi2007}. In terms of rms value, the GH fit (rms=6.76) is a better model than the Lundquist model (rms=9.49) as compared to the observations in Figure~\ref{fig_mc}. Based on these fits, the MC axis is oriented -15\degree~in latitude and roughly 260\degree~in longitude. These results indicate that the MC magnetic fields had a significant southward Bz component to launch the geomagnetic storm. The flux rope has a positive sign of magnetic helicity which is inline with the source region helicity sign of the erupting part of the filament as seen in EUV images (Figure~\ref{fig_aia_fil}). The twist of the magnetic field in the GH flux rope model is 1.67 turns/au. The MC has a diameter of 0.17 au.  While passing the spacecraft, the radius of the MC varies with time: $r(t) = r_0(1 + Et)$ where E is the expansion rate determined to be 0.09 per day from the MC fitting analysis.
 
\begin{table}[!ht]
	\centering	
	\caption{Chronology of the CME eruption from AR 13283}
	\begin{tabular}{ll}
 \hline
    21/17:44 UT     & CME initiation \& flare start time \\
    21/18:03 UT     & Onset of CME eruption \\
    21/18:12 UT     & peak time of M1.7 flare  \\
    21/20:30 UT    & CME LE at a height 20 R$_\odot$  \\   
    23/07:30 UT     & ICME shock arrival \\
    24/01:00-15:00 UT    & MC passage \\
    23/08:30 UT    & geomagnetic storm commencement \\
    24/05:30 UT    & geomagnetic storm peaked at 212 nT \\
    \hline
\end{tabular}
\label{tab:Tab1}
\end{table}

\section{Summary and Discussion}
\label{SummDisc}
We investigated the CME eruption that occurred on April 21, 2023 which caused a severe geomagnetic storm on April 23, 2023. The eruption originated from solar AR 13283 near the disk center. The AR was in its decay stage, with fragmented polarities and a pre-existing long filament channel, a few days before the eruption. The polarities have undergone large-scale shear and converging motions, so that the flux cancellation leads to the buildup of twisted flux threads along the main PIL. The magnetic helicity is positive, so the chirality of the filament threads is right-handed. Importantly, the helicity injection changed from a positive to a negative sign early on April 21. Because the flux rope (filament) has been built up by monotonous helicity accumulation over several days \citep{amari2003b, vemareddy2015_FullStudy_SunEarth}, the further converging and canceling fluxes lead to helicity injection change and the unstable nature of the MFR and its further eruption. 

The filament that is initiated to erupt is a part of the long, multiply-threaded filament channel. Along with the increase in GOES X-ray flux, the filament set to rise at 17:42 UT and eventually erupted at around 18:03 UT (see Table~\ref{tab:Tab1}). Importantly, the CME morphology revealed that the MFR is oriented in the north-east to south-west direction (-26\degree~with respect to the equator), which differs by 56\degree with respect to the filament orientation, which is in the south-east to north-west direction (30\degree~with respect to the equator). These observations imply that the filament apex rotates in a clockwise direction owing to the right-hand twist of the magnetic field \citep{lynch2009, green2011}.

The CME decelerates in the LASCO fov and has a velocity of 1226 km/s at 20 R$_\odot$. In the HI fov, the CME lateral expansion is tracked more than the earthward motion, then the arrival time estimation is difficult to assess. However, the drag-based model gives correct estimates of the arrival time of the ICME sheath with the initial CME speed at 20 R$_\odot$. The in situ arrival of the ICME shock was at 07:30 UT on April 23, and then a geomagnetic storm commenced at 08:30 UT. The flux rope fitting to the in situ magnetic field observations reveal that the MC flux rope orientation is consistent with its near Sun orientation within 10 R$_\odot$, which has a significant negative Bz-component. The analysis of this study indicates that the near-Sun rotation of the filament during its eruption to the CME is the key to the negative $B_z$-component and consequently the intense geomagnetic storm.  \\


\begin{acknowledgements} SDO is a mission of NASA’s Living With a Star Program; STEREO is the third mission in NASA’s Solar Terrestrial Probes program; and SOHO is a mission of international cooperation between the ESA and NASA. We acknowledge the use of NASA/GSFC's Space Physics Data Facility's OMNIWeb (or CDAWeb or ftp) service. The CME catalog used in this study is generated and maintained by the Center for Solar Physics and Space Weather, the Catholic University of America, in cooperation with the Naval Research Laboratory and NASA. The author is grateful to the anonymous referee for a detailed list of comments and suggestions.
\end{acknowledgements}
\bibliographystyle{aasjournal}

\end{document}